\documentclass[floatfix,aps,twocolumn,superscriptaddress]{revtex4-1}

\usepackage{amssymb}
\usepackage{amsmath}
\usepackage[dvips]{graphicx}
\usepackage{bm}
\usepackage{float}
\usepackage{color}

\newcommand{\be}{\begin{equation}}
\newcommand{\ee}{\end{equation}}
\newcommand{\bea}{\begin{eqnarray}}
\newcommand{\eea}{\end{eqnarray}}

\begin{document}

\title{Ferromagnetic and underscreened Kondo behavior in quantum dot arrays}

\author{J. A. Andrade}
\affiliation{Centro At{\'o}mico Bariloche and Instituto Balseiro, CNEA, 8400 Bariloche, Argentina}
\affiliation{Consejo Nacional de Investigaciones Cient\'{\i}ficas y T\'ecnicas (CONICET), Argentina}
\author{D. J. Garc\'ia}
\affiliation{Centro At{\'o}mico Bariloche and Instituto Balseiro, CNEA, 8400 Bariloche, Argentina}
\affiliation{Consejo Nacional de Investigaciones Cient\'{\i}ficas y T\'ecnicas (CONICET), Argentina}
\author{Pablo S. Cornaglia}
\affiliation{Centro At{\'o}mico Bariloche and Instituto Balseiro, CNEA, 8400 Bariloche, Argentina}
\affiliation{Consejo Nacional de Investigaciones Cient\'{\i}ficas y T\'ecnicas (CONICET), Argentina}

\begin{abstract}
    We analyze the low energy properties of a device with $N+1$ quantum dots in a star configuration. A central quantum dot is tunnel coupled to source and drain electrodes and to $N$ quantum dots. Extending previous results for the $N=2$ case we show that, in the appropriate parameter regime, the low energy Hamiltonian of the system is a ferromagnetic Kondo model for a $S=(N-1)/2$ impurity spin. For small enough interdot tunnel coupling, however, a two-stage Kondo effect takes place as the temperature is decreased. The spin $1/2$ in the central quantum dot is Kondo screened first and at lower temperatures the antiferromagnetic coupling to the side-coupled quantum dots leads to an underscreened $S=N/2$ Kondo effect. 
We present numerical results for the thermodynamic and spectral properties of the system which show a singular behavior at low temperatures and allow to characterize the different strongly correlated regimes of the device. 
\end{abstract}


\maketitle
\section{Introduction}
The ferromagnetic and underscreened Kondo models are examples of quantum impurity problems where the low energy properties cannot be described using Fermi-liquid theory \cite{nozieres1974fermi,pines1990theory}. 
In these models, a local magnetic moment decouples asymptotically at low energies from a non-interacting electron reservoir leading to a singular behavior in the thermodynamic, dynamic, and transport properties of the impurity \cite{nozieres1980kondo,coleman2003pepin,mehta2005regular,koller2005singular,cornaglia2011quantum,florens2011universal,cabrera2013magneto}. The resulting asymptotically free magnetic moment is extremely sensitive to a Zeeman splitting and at zero temperature is easily polarized by any non-zero external magnetic field. The lifting of the degeneracy by the magnetic field changes the nature of the ground state driving the system into a Fermi liquid regime.
This extreme sensitivity to external fields could be used in nanoscopic devices to control the electronic and thermal transport and to generate spin currents \cite{Cornaglia2012}.

Underscreened Kondo physics has been recently reported in spin-1 molecular junctions and multilevel quantum dots where a single channel (not counting spin) in the electrodes is relevant for the screening process for a wide range of temperatures \cite{sasaki2000kondo,roch2008quantum,roch2009observation,Parks11062010}. In these systems, however, the magnetic moment is expected to be fully screened at low enough temperatures since the coupling to a second conduction channel cannot be ruled out by symmetry considerations. Quantum dot (QD) devices in semiconductor heterostructures offer the possibility of tuning the parameters and setting the geometry to obtain different quantum impurity models \cite{camjayi2012kondodis}.
Devices having three QDs in a star configuration have been shown to lead to the spin-$1/2$ ferromagnetic Kondo model \cite{kuzmenko2006,mitchell2009,mitchell2013,Baruselli2013}. The realization of such devices would allow the first unambiguous measurement of ferromagnetic Kondo physics. 
Here we show that including additional side-coupled QDs the system can be used to construct, by appropiatedly setting the parameters, the ferromagnetic Kondo model for a spin $S= (N-1)/2$, where $N$ is the number of side-coupled QDs, and the underscreened Kondo model for a spin $S=N/2$.
\begin{figure}[t]
\includegraphics[width=0.8\columnwidth]{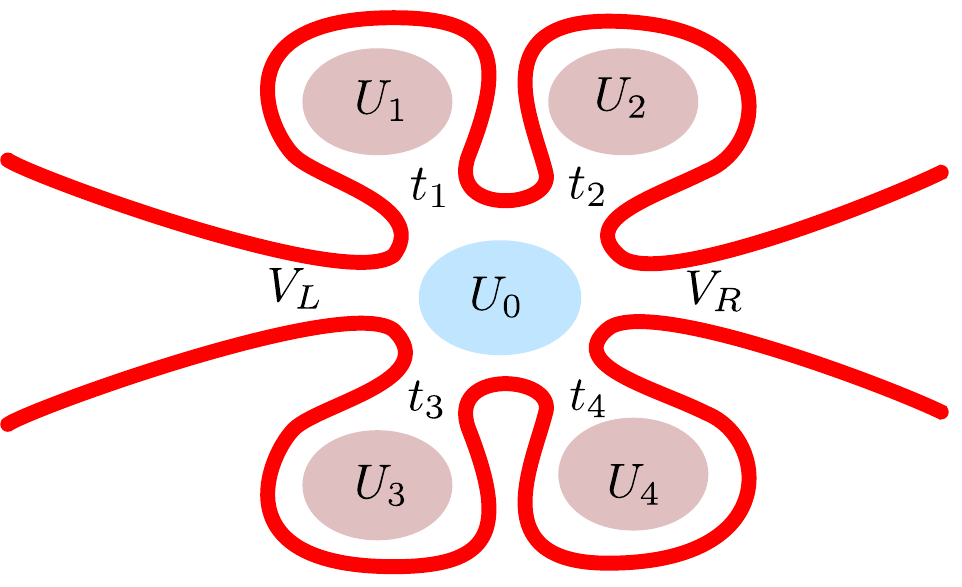}
\caption{(Color online) Schematic representation of a $5$ QD device in a star configuration. The charging energies $U_\ell$ and hopping matrix elements $t_\ell$ for each QD are indicated in the figure. $V_L$ and $V_R$ are the tunnel couplings to left and right electrodes, respectively.}
\label{fig:device}
\end{figure}

In the weak QD-electrodes coupling regime, the low energy properties of the device are dominated by the ground state manifold of the isolated QD array which, as we show below, has spin $S=(N-1)/2$ and is ferromagnetically coupled to the electrodes. 
In the opposite regime, the magnetic moment of the central QD is screened through a regular Kondo effect leading to a local Fermi liquid of quasiparticles \cite{nozieres1974fermi}. The side-coupled QDs form a spin $S=N/2$ which is antiferromagnetically coupled to the central QD. To fully quench the associated magnetic moment through Kondo screening, $2S$ conduction electron channels would be needed. As we show below, there is a single conduction channel available and the magnetic moment is only partially screened. 
In both regimes and in the crossover between them, there is an asymptotically free $S=(N-1)/2$ local magnetic moment at low energies.

We calculate numerically the properties of devices with up to 6 QDs and obtain the expected logarithmic behavior in the thermodynamic and dynamic properties in the ferromagnetic and underscreened Kondo regimes and in the crossover between them.

The rest of the paper is organized as follows: In Sec. \ref{sec:model} we present the model for the $N+1$ QD device. In Sec. \ref{sec:lowE} we show that in the appropriate parameter regime, the low energy properties can be described by a $S=(N-1)/2$ ferromagnetic Kondo model. Section \ref{sec:numer} presents the numerical results, using the density matrix extension of Wilson's numerical renormalization group (DM-NRG), for the spectral density of the central QD and the magnetic susceptibility.
Finally, in Sec. \ref{sec:concl} we present our conclusions.

\section{Model}\label{sec:model}
We consider $N+1$ QDs, with a single relevant electronic level on each QD \footnote{For a small enough quantum dot, the transport properties are governed by its lowest unoccupied or the highest occupied orbital.}, coupled in a star configuration (see Fig. \ref{fig:device}). The device is described by the following Hamiltonian \cite{Baruselli2013}

\begin{equation} \label{eq:hamilt}
  H=H_C+ H_{t}+H_{V}+H_{\text el}\;.
\end{equation}

Here $H_C$ describes the electrostatic interaction on the QDs 
\begin{equation} \label{eq:coul}
H_C = \sum_{\ell=0}^{N}\left(\varepsilon_{\ell}\sum_{\sigma=\uparrow,\downarrow}\hat{n}_{\ell\sigma}+ U_\ell \hat{n}_{\ell\uparrow}\hat{n}_{\ell\downarrow}\right),
\end{equation}
where  $\hat{n}_{\ell\sigma}= d_{\ell\sigma}^\dagger d_{\ell\sigma}$ is the electron number operator of the $\ell$-th QD, $U_\ell>0$ is its charging energy, and $\sigma$ is the electron spin projection along the $\hat{z}$ axis. The level energy $\varepsilon_{\ell}$ can be controlled experimentally via a gate voltage $V_{g\ell}$, $\varepsilon_\ell \simeq -C_{g\ell} V_{g\ell}$, where $C_{g\ell}$ is the capacitance of QD $\ell$ with its corresponding gate electrode. 
\begin{equation} \label{eq:hopp}
H_t = \sum_{\sigma}\sum_{\ell=1}^{N}\left(t_{\ell}d_{\ell\sigma}^{\dagger}d_{0\sigma}+h.c.\right),
\end{equation}
describes the tunneling coupling between the central QD ($\ell=0$) and the $N$ side-coupled QDs. Finally,

\begin{equation} \label{eq:hybr}
    H_{V} = \sum_{\nu=L,R}\sum_{k,\sigma}\left(V_{k\nu}d_{0\sigma}^{\dagger}c_{\nu k\sigma}+h.c.\right),
\end{equation}
describes the coupling between the central QD and the left ($L$) and right ($R$) electrodes, 
which are modeled by two non-interacting Fermi gases:

\begin{equation}
H_{\text el} = \sum_{\nu, k,\sigma} \epsilon_k c_{\nu k\sigma}^\dagger c_{\nu k\sigma}.
\end{equation}

\section{Low energy Hamiltonians}\label{sec:lowE}
\subsection{Decoupled impurity}
We first analyze the nature of the eigenstates of $H_{QD}=H_C+H_t$ which describes the system of QDs decoupled from the electrodes. We focus on the regime with a single electron on average on each QD and perform a Schrieffer-Wolff transformation (valid for: $\varepsilon_\ell<0$, $\varepsilon_\ell+U_\ell>0$, and $t_{\ell}\ll \min(|\varepsilon_\ell|,\varepsilon_\ell +U_\ell$)) to decouple the high energy empty and double occupied states on each QD.

Up to second order on the tunnel coupling term $H_t$ the Hamiltonian reads $H_{QD}=H_J+H_W+H_C$. Here $H_W$ describes density-density interactions between the central and the side-coupled QDs \cite{Schrieffer-Wolff} and $H_J$ describes the exchange interaction between the central QD and the side-coupled QDs:
\begin{equation} 
    H_J = \sum_{\ell=1}^{N} J_{\ell}\mathbf{S_{0}\cdot S_{\ell}},
\label{MultiHeisenb}
\end{equation}
where $\mathbf{S_\ell}=\sum_{\sigma \sigma^\prime} d^\dagger_{\ell}\frac{\boldsymbol{\sigma_{\sigma \sigma^\prime}}}{2}d_{\ell \sigma^\prime}$ and $\boldsymbol{\sigma}$ is a vector composed by the Pauli matrices $\boldsymbol{\sigma}=(\sigma_x,\sigma_y,\sigma_z)$.
The antiferromagnetic exchange couplings $J_\ell>0$ are given by
\begin{equation} \label{eq:coupl}
J_{\ell} = 8t_{\ell}^{2}\left(\frac{U_0+U_\ell}{(U_0+U_{\ell})^2-4(\varepsilon_{0}-\varepsilon_{\ell})^2}\right).
\end{equation}
For the spin Hamiltonian of Eq. (\ref{MultiHeisenb}) two sublattices can be defined with no interaction between spins on the same sublattice, one formed by the central QD (sublattice A) and other formed by the side-coupled QDs (sublattice B). Under these conditions, the Lieb-Mattis theorem \cite{lieb1962ordering} states that the ground state of the system has total spin $S=|S_A-S_B|$, where $S_A$ ($S_B$) is the maximum possible total spin on sublattice $A$ ($B$). This implies that the ground state of the isolated QD system will have total spin $S=(N-1)/2$. 

If we consider identical couplings $J_\ell=J$ between the QDs, the spin Hamiltonian reduces to an antiferromagnetic exchange coupling $H_J\to J \mathbf{S_{0}\cdot S_D}$  between the spin on the central QD and the total spin of the side-coupled QDs $\mathbf{S_{D}}=\sum_{\ell=1}^{N} \mathbf{S_{\ell}}$ and can be readily solved.
The eigenstates of $H_J$ can be obtained adding the angular momenta of the two sublattices  $\mathbf{L}= \mathbf{S_0} + \mathbf{S_D}$. 
Using $J \mathbf{S_{0}\cdot S_D} =J(L^2-S_D^2-S_0^2)/2$ we obtain the corresponding eigenvalues:
\begin{eqnarray} \label{eq:eigval}
    E_{L=S_D\pm \frac{1}{2}}&=& -\tfrac{J}{2}(\mp(S_D+1)+\tfrac{1}{2}).
\end{eqnarray}
The minimum energy is obtained for the maximum possible $S_D=N/2$ which results in $L=(N-1)/2$ as expected from Lieb-Mattis theorem. The corresponding  eigenstate is 
\begin{eqnarray} \label{ground-state}
    |L,L_m\rangle &=& \frac{1}{\sqrt{2S_D+1}}\left(\sqrt{S_D+L_m+\tfrac{1}{2}}\mid \downarrow\rangle|S_D,L_m+\tfrac{1}{2}\rangle\right.\nonumber\\
    &-& \left.\sqrt{S_D-L_m+\tfrac{1}{2}}\mid \uparrow\rangle\mid S_D,L_m-\tfrac{1}{2}\rangle\right),
\end{eqnarray}
where $S_D=N/2$, $|\sigma\rangle$ and $|S_D,S_z \rangle$ are the eigenstates of $S_0^2$ and $S_D^2$, respectively, with projections $\sigma$ and $S_z$ along the $\hat{z}$ axis, and $L_m$ is the projection of the total angular momentum along the same axis. 

\subsection{Ferromagnetic Kondo Hamiltonian}
In this Section we include the coupling to the electrodes and show that, in the appropriate parameter regime, the ferromagnetic Kondo model for an impurity spin $S=(N-1)/2$ describes the low energy spin dynamics of the $N+1$ QD device. 
To that aim it is convenient to rewrite the coupling between the central QD and the electrodes [see Eq. (\ref{eq:hybr})] as 
\begin{equation}
    H_V =  V \sum_\sigma \left( d_{0\sigma}^\dagger c_{1\sigma} + h.c. \right), 
    \label{eq:c0}
\end{equation}
where $V=\sqrt{\sum_{k\nu} |V_{k\nu}|^2}$ and $c_{1\sigma}=\tfrac{1}{V}\sum_{k \nu}V_{k\nu}c_{k\nu\sigma}$ destroys an electron on the state $|\Psi_1\rangle$ of the electrodes to which the central QD is coupled. 

To calculate the ferromagnetic coupling, we perform a second-order perturbation theory in $H_V$ (Schrieffer-Wolff transformation) neglecting potential scattering terms and retaining only the magnetic coupling between the band electrons and the ground state manifold of the isolated QD array [see  Eq. (\ref{ground-state})].  Following the notation and procedure of Ref. \cite{Baruselli2013} for $J_\ell=J$ we obtain an exchange coupling which is given by:
\begin{eqnarray} \label{Jk_N_Lm}
	J_{K} &=& \frac{\langle L,L_m+1|d_{0\uparrow}^{\dagger}\frac{1}{H_{QD}-E_g} d_{0\downarrow}- d_{0\downarrow}\frac{1}{H_{QD}-E_g} d_{0\uparrow}^{\dagger}|L,L_m\rangle}{\sqrt{(L+\tfrac{1}{2})^2-(L_m+\tfrac{1}{2})^2}},\nonumber
\end{eqnarray}
where $L=S_D-1/2=(N-1)/2$, $E_g=E_{L=S_D-1/2}$ and we have neglected the energy of the conduction electrons, compared to the charging energies, since we are focussing our analysis on the low energy properties of the system.
We finally obtain a ferromagnetic coupling
\begin{equation} \label{Jk_N}
J_K=-\frac{16V^2}{(N+1)(J(N+2)+2U_0)},
\end{equation}
for the Kondo Hamiltonian 
\begin{equation} 
	H_K =  J_K\mathbf{L}\cdot \mathbf{s} + H_{el},
\label{eq:Kondoham}
\end{equation}
where $\mathbf{s}=\sum_{\sigma \sigma^\prime} c^\dagger_{1\sigma}\frac{\boldsymbol{\sigma_{\sigma \sigma^\prime}}}{2}c_{1 \sigma^\prime}$ is the spin operator of the local state of the electrodes.
This result, which  extends the $N=2$ analysis of Baruselli {\it el al.} to arbitrary $N$, assumes that the ground state manifold of the isolated QD is the only relevant at low energies. As we will see below this is not the case if the coupling to the electrodes is strong enough.

To show this, we perform the Schrieffer-Wolff transformation taking $H_V$ as a perturbation without projecting to the ground state manifold of the isolated QD array. We obtain the following Hamiltonian:
\begin{equation}
	H_1 = \sum_{\ell=1}^{N} J_{\ell}\mathbf{S_{0}\cdot S_{\ell}} + J_V \mathbf{S_{0}\cdot \mathbf{s}} +H_{el}, 
	\label{eq:Heis2}
\end{equation}
where $J_V=8V^2/U_0$ is an antiferromagnetic coupling. To arrive to Eq. (\ref{eq:Heis2}) we have discarded potential scattering terms for simplicity, and we have assumed $N J_\ell\ll U_0$.
In the case that $J_\ell=J\gg J_V$, we expect the low energy physics to be dominated by the ground state of the isolated QD array and we project $H_1$ into its manifold: $H_1\to \langle m| H_1 |m^\prime \rangle $ where $|m\rangle\equiv|L,L_m\rangle$.
After the projection, the first term of $H_1$ turns into a constant that we discard. The projection of the second term can be evaluated using the Wigner-Eckart theorem based result \cite{merzbacher1998quantum}:
\begin{equation}
	\langle m|\mathbf{S_{0}}|  m^\prime\rangle=\frac{\langle m|\mathbf{S_{0}\cdot \mathbf{L}}|  m\rangle}{L(L+1)}\langle m|\mathbf{L}|  m^\prime\rangle,
	\label{eq:proj2}
\end{equation}
which for the states of Eq. (\ref{ground-state}) is:
\begin{equation}
	\langle m|\mathbf{S_{0}}|  m^\prime\rangle=-\frac{1}{N+1}\langle m|\mathbf{L}|  m^\prime\rangle.
	\label{eq:proj3}
\end{equation}
The minus sign of the proportionality coefficient between $\mathbf{S_0}$ and $\mathbf{L}$ leads to a ferromagnetic coupling $J_K=-J_V/(N+1)$ between $\mathbf{s}$ and $\mathbf{L}$ out of an antiferromagnetic coupling between  $\mathbf{s}$ and $\mathbf{S_0}$, and to recover $H_K$ as written in Eq. (\ref{eq:Kondoham}). Note however that to calculate $J_V$ we have assumed $N J \ll U_0$, if this is not the case we obtain $J_V =\frac{16V^2}{(N+2)J + 2U_0}$ and recover the Kondo coupling of Eq. (\ref{Jk_N}) for the case $J_\ell=J$.

In the general situation where the $J_\ell$ are not all identical, the Hamiltonian $H_J$ no longer commutes with $S_D^2$. The ground state of the isolated QD array [which is described by Eq. (\ref{MultiHeisenb})] is, for a given $L_m$, a superposition of the $N$ states that have total spin angular momentum $L=(N-1)/2$, of which a single one has $S_D=N/2$ while the remaining $N-1$ states have $S_D=N/2-1$. It can therefore be written as:
\begin{equation}
	|L\rangle_{GS} = \cos\theta |L,S_D=\tfrac{N}{2}\rangle +  \sin\theta |L,S_D=\tfrac{N}{2}-1\rangle
\label{eq:genGS}
\end{equation}
where $\theta$ is a real number.
Projecting $H_1$ into the ground state manifold we obtain a Kondo coupling:
\begin{equation}
	J_K=\left(-\frac{\cos^2\theta}{N+1}+\  \frac{\sin^2\theta}{N-1}\right)J_V,
	\label{eq:genJk}
\end{equation}
which, as we show in the following section is always ferromagnetic and attains its maximum absolute value for uniform couplings ($J_\ell=J$). The minimum absolute value of $J_K$ is zero and is obtained in the limit where one of the $J_\ell$ couplings dominates over the others (e.g. $J_{\ell\neq 1}/J_1\to 0$). 
\subsection{Proof of $J_K<0$}\label{sec:proof}
Here we show that the Kondo coupling of Eq. (\ref{eq:genJk}) is negative for arbitrary values of $N$ and $J_\ell>0$. 
To calculate $J_K$ from Eq. (\ref{eq:genJk}) only the coefficient $\theta$ from the ground state wave-function of the isolated QD array [see Eq. (\ref{eq:genGS})] needs to be obtained. It can be calculated diagonalizing the $N\times N$ Hamiltonian matrix associated with the ground state subspace, which we managed to do analytically for arbitrary $N$ only in a few highly symmetric cases.    
To show that the coupling is ferromagnetic, however, it suffices to obtain a lower bound on the value of $\cos^2\theta$. To that aim we propose a ground state wave-function of the form:
\begin{equation}
	|\psi\rangle= \sum_i\alpha_i |S_i\rangle,
	\label{eq:gswf}
\end{equation}
where the $\alpha_i$ are real numbers and the 
\begin{equation}
    |S_i\rangle = \frac{1}{\sqrt{2}}\left(|\downarrow\uparrow\cdots\uparrow\rangle-|\uparrow\cdots\uparrow\underset{i}{\downarrow}\uparrow\cdots\uparrow\rangle\right),
	\label{eq:singlet}
\end{equation}
are ``singlets'' between the central QD and the $i$-th QD, with the rest of the spins parallel. Here, the ket $|a_0a_1\cdots a_N\rangle$ is a state in the Fock basis where the spin projections along the z-axis for the spin-1/2 on each QD are given by the $a_\ell$. The $|S_i\rangle$ form a non-orthogonal basis of the subspace with total angular momentum $L=(N-1)/2$ where the ground state is to be found according to Lieb-Mattis theorem. This makes a variational calculation of the $\alpha_i$ exact for the ground state wave-function. 
Applying $H_J$ to the proposed wave-function the following linear combination of the $|S_i\rangle$ is obtained
\begin{equation}
	H_J|\Psi\rangle = \sum_i \left\{-\frac{3}{4} \alpha_i J_i +\frac{1}{2}\sum_{j\neq i} \left(\alpha_i \frac{J_j}{2} -\alpha_jJ_i\right) \right\}|S_i\rangle,
\end{equation}
 from which the variational energy $E_{var}=\langle\Psi|H_J|\Psi\rangle$ can be calculated
\begin{equation}
	E_{var}=-\frac{3}{4}\sum_i J_i \left\{\sum_{j}\alpha_i \alpha_j +\frac{1}{6} \sum_{j\neq i, k\neq j, k\neq i } \alpha_j \alpha_k   \right\}, 
	\label{evar}
\end{equation}
that has to be minimized under the normalization constraint: 
\begin{equation}
	\langle \Psi|\Psi\rangle=\frac{1}{2}\left(\sum_i \alpha_i\right)^2 + \frac{1}{2}\sum_{i}\alpha_i^2 =1.
    \label{eq:norm}
\end{equation}

Since all the coefficients in the quadratic form of Eq. (\ref{evar}) have the same negative sign, the $\alpha_i$ that minimize it will also have the same sign and can be chosen as positive ($\alpha_i>0$) without loss of generality.
The coefficient $\theta$ can be calculated from the projection of $|\Psi\rangle$ on the state 
\begin{equation}
	|L_z=L,S_D=N/2\rangle=\sqrt{\frac{2}{N(N+1)}} \sum_i|S_i\rangle,
\end{equation}
which leads to
\begin{equation}
	\cos \theta = \langle\Psi	|L_z=L,S_D=N/2\rangle=\sqrt{\frac{N+1}{2N}}\sum_i \alpha_i.
\end{equation} 
Replacing the calculated $\cos\theta$ in Eq. (\ref{eq:genJk}) results in
\begin{equation}
	J_K=\frac{J_V}{N-1}\left[1-\left( \sum_i \alpha_i  \right)^2 \right].
	\label{JKvar}
\end{equation}
Due to the positiveness of the $\alpha_i$ we have
\begin{equation}
	\left( \sum_i \alpha_i \right)^2=\sum_i \alpha_i^2+ \sum_i \sum_{j\neq i}\alpha_i \alpha_j\geq 	\sum_i \alpha_i^2,	
\end{equation}
which using the normalization condition of Eq. (\ref{eq:norm}) leads to $\sum_i\alpha_i^2\leq 1$ and
\begin{equation}
	\left( \sum_i \alpha_i \right)^2\geq 1.
    \label{eq:ineq}
\end{equation}

Combining Eq. (\ref{eq:ineq}) and Eq. (\ref{JKvar}) we finally obtain $J_K\leq 0$, showing that the coupling is ferromagnetic.
When one of the couplings dominates over the others $J_{\ell\neq 1} /J_1\to 0$ we have $\alpha_1=1$, $\alpha_{\ell\neq 1}=0$ and $J_K=0$. The maximum $J_K$ in absolute value is obtained for symmetric couplings $J_\ell=J$ for which $\alpha_\ell=\sqrt{\frac{2}{N(N+1)}}$ and $J_K=-J_V/(N+1)$.

\subsection{Two-stage Kondo regime}
The projection of $H_1$ to the ground state manifold of the isolated QDs is justified if the energy gap between the ground state and the first excited state is much larger than the coupling energy to the electrodes. 
If we assume the opposite situation $J_V\gg J_\ell$, the low energy physics can no longer be described by the ferromagnetic Kondo Hamiltonian. For $J_\ell \to 0$, the model reduces to the usual spin $1/2$ Kondo Hamiltonian and the local magnetic moment of the central QD is screened at temperatures below the Kondo temperature \cite{wilson1975kondo}
\begin{equation}
    T_K^0 = \frac{D}{k_B} \sqrt{\rho_0 J_V} \exp[-1/\rho_0 J_V],
	\label{ieq:tk0}
\end{equation}
where $D$ is a high energy cutoff, and $\rho_0$ is the local density of states of the electrodes at the effective state $|\Psi_1\rangle$ evaluated at the Fermi energy. 
The local density of states at the central QD presents an Abrikosov-Suhl resonance of width $\sim k_B T_K^0$ and the local low energy properties can be described by a Fermi liquid with quasiparticles having a renormalized mass $\propto 1/T_K^0$ \cite{nozieres1974fermi}.

If we turn on a small $J_\ell=J\ll k_B T_K^0$, the local spins on the side-coupled QDs couple antiferromagnetically to the local renormalized Fermi liquid at the central QD.  This situation has been analyzed for $N=1$ using NRG and a slave-boson mean field approach in Ref. \cite{cornaglia2005strongly}. The main conclusion is that for $J<k_B T_K^0$ there is a second Kondo effect with a characteristic temperature:
\begin{equation}
    T_K^\star\sim T_K^0 \exp[-\pi k_B T_K^0/J],
    \label{eq:tksecond}
\end{equation}
which corresponds to a Kondo effect due to the antiferromagnetic coupling $J$ between a spin-1/2 and a Fermi liquid having an effective bandwidth $\widetilde{D}\sim k_B T_K^0$ and a local density of states at the Fermi level $\widetilde{\rho}_0\sim 1/\pi k_B T_K^0$. 

As we show below, for $N>1$ the situation is analogous  to the $N=1$ case in the $J\ll T_K^0$ regime. The main difference being that for $N=1$ the spin on the side-coupled QD is fully screened while for $N>1$ the conduction electron channel available at the central QD is unable to fully screen the local magnetic moments at the side coupled QDs, leading to an underscreened Kondo effect. A numerical analysis shows that the characteristic temperature of the underscreened Kondo effect is well described by Eq. (\ref{eq:tksecond}).   

\section{Numerical results} \label{sec:numer}
We calculate the thermodynamic and spectral properties of the QD system using Wilson's Numerical Renormalization Group \cite{wilson1975renormalization,krishna1980renormalizationI,krishna1980renormalizationII,bulla2008numerical,campo2005alternative,oliveira1994generalized,yoshida1990renormalization} with the density matrix extension \cite{Andreas2007} and z-trick \cite{Oliveira1990} to improve the accuracy of the low temperatures properties and the high energy resolution of the spectral densities, respectively. The electron bath provided by the electrodes is characterized by the hybridization $\Delta(\omega)=\pi \rho_0(\omega) V^2$ where $\rho_0(\omega)$ is the local density of states of the electrodes at the orbital $|\psi_1\rangle$ to which the central QD is coupled.  For simplicity we take  $\Delta$ as energy independent and its support within the range $[-D,D]$, where $D$ is the half bandwidth of the conduction electron band. In what folows we select energy units such that $D=1$ and take the Fermi energy at $\epsilon_F=0$.
\subsection{Ferromagnetic to underscreened crossover ($N=2$)}
\begin{figure}[tbp]
\includegraphics[width=\columnwidth]{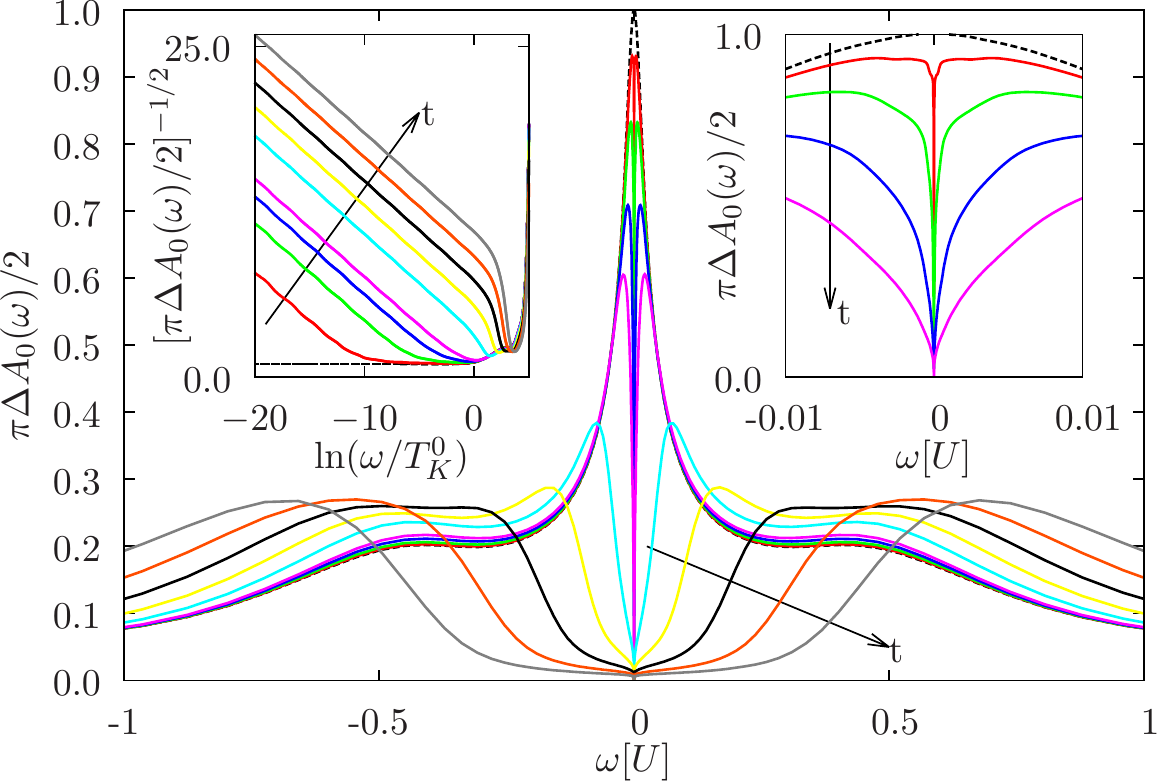}
\caption{(Color online) Spectral density of the central QD of a device with two side-coupled QDs ($N=2$). The parameters are $U_\ell=U=0.4$, $\varepsilon_\ell=-U/2$, $\Delta=0.04\pi$ and $t_\ell=t=0$, $0.015$, $0.02$, $0.025$, $0.03$, $0.05$, $0.07$, $0.09$, $0.11$, and $0.13$. Left inset: plot of $A_0(\omega)^{-1/2}$ as a function of $\ln(\omega)$ to make clear the singular behavior of the spectral density at low energies for $t\neq 0$. Right inset: low energy detail of the spectral density for $t_\ell=t=0$, $0.015$, $0.02$, $0.025$, and $0.03$.}
\label{Aw-2QD-tx}
\end{figure}
Figure \ref{Aw-2QD-tx} presents the zero-temperature spectral density of the central QD  $A_{0}(\omega,T=0)$ in a $N=2$ device, with identical QDs ($U_\ell=U$, $\varepsilon_\ell=-U/2$), for different values of the interdot tunnel coupling $t_\ell=t$. In this parameter regime there is a single electron on average on each QD. For $t=0$ we obtain an Abrikosov-Suhl resonance at the Fermi level associated with the screening of the magnetic moment of the central QD. The full width at half maximum of the resonance is $\sim k_BT_K^0$ and its height at the Fermi level $A_0(\omega=0)=2/\pi\Delta$ according to Friedel's sum rule \cite{langreth1966friedel} for the Anderson model. There are also two peaks at $\omega \simeq \varepsilon_0$, and $U_0+\varepsilon_0$ associated with charge fluctuations.

When a small coupling $t$ between the QDs is included the spectral density develops a dip and vanishes at the Fermi level. The dip is associated with the underscreening of the magnetic moment of the side-coupled QDs and leads to a complete suppression of the spectral density at the Fermi level at zero temperature. This dip can be interpreted in terms of a Kondo hole formed as a consequence of the strong coupling between the side-coupled QDs and the central QD.
For $N=2$ the vanishing of the spectral density at the Fermi level is contrary to the expectations from Fermi liquid theory and Luttinger's theorem \cite{abrikosov1975methods} that predicts for this system in the wide-bandwith limit \cite{andrade2014transport}
\begin{equation}
	A^{FL}_0(\omega=0,T=0) = \sum_\sigma \frac{\sin^2(\mathcal{N}_{\sigma}\pi)}{\pi\Delta},
	\label{eq:FL}
\end{equation}
where $\mathcal{N}_\sigma$ is the total occupancy per spin of the QD array \footnote{More precisely, the displaced charge. See Ref. \cite{langreth1966friedel}}. In the absence of a magnetic field we have $\mathcal{N}_\uparrow=\mathcal{N}_\downarrow$ and in the electron-hole symmetric situation we have $\mathcal{N_\sigma}=(N+1)/2$, and 

\begin{equation}
	A^{FL}_0(\omega=0,T=0) = \left\{\begin{matrix} 0\,\,\,&\text{for odd } N\\ 2/\pi\Delta\,\,\, &\text{for even } N\end{matrix}\right..
	\label{eq:AFL}
\end{equation}
For $t>0$ the spectral densities of Fig. \ref{Aw-2QD-tx} do not satisfy Eq. (\ref{eq:AFL}) indicating a non-Fermi liquid (NFL) behavior and a non-vanishing Luttinger integral \cite{logan2014common}.
In the left inset to Fig. \ref{Aw-2QD-tx} we plot $[A_0(\omega)]^{-1/2}$ as a function of $\ln(\omega)$ which shows a linear behavior at low energies consistent with a singular behavior of the form 
\begin{equation}
	A_0(\omega\to 0)\simeq  \frac{b}{\ln^2(|\omega|/k_B T_0)},
	\label{singAlow}
\end{equation}
at low energies, as expected for the ferromagnetic and underscreened Kondo models \cite{koller2005singular,logan2009correlated}.
The width of the dip increases with increasing $t$ (see right inset to Fig. \ref{Aw-2QD-tx}) and suppresses the Abrikosov-Suhl resonance. For large enough $t$ the Kondo effect associated with the screening of the central QD does not develop and the system is in the ferromagnetic Kondo regime. 
\begin{figure}[tbp]
\includegraphics[width=\columnwidth]{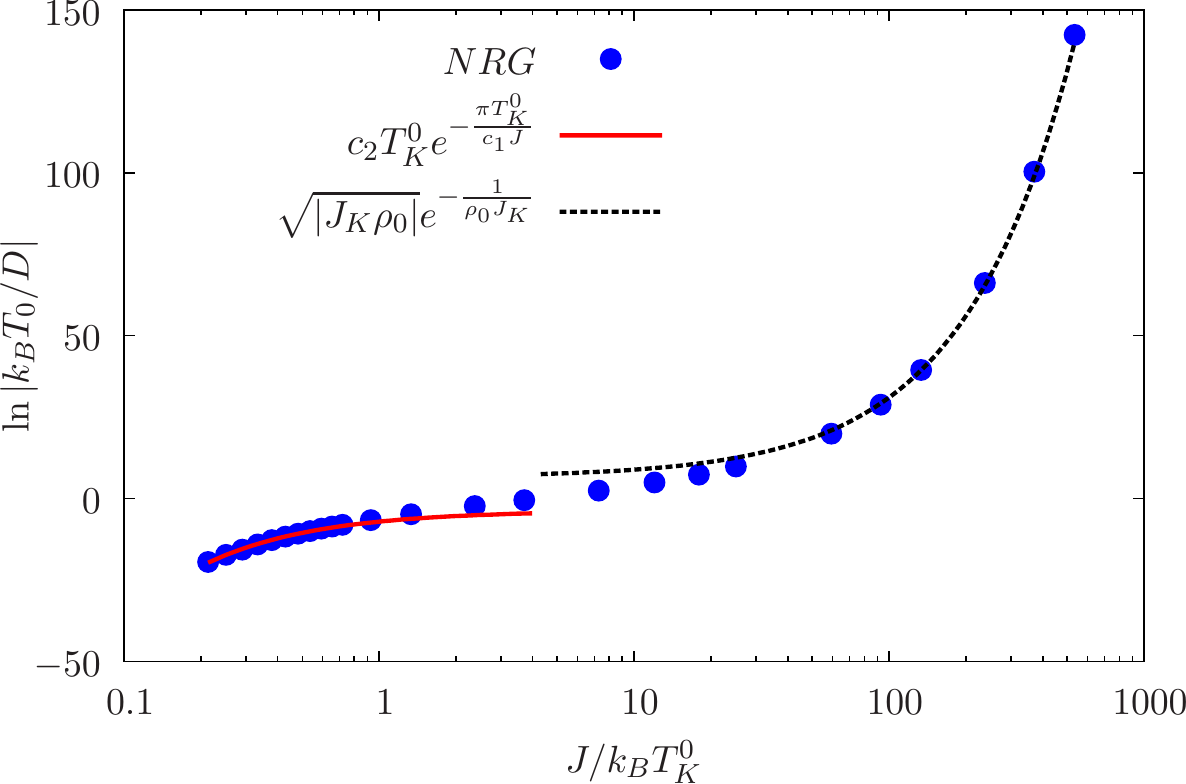}
\caption{(Color online) Behavior of the low energy scale $k_BT_0$ as a function of the interdot antiferromagnetic coupling $J=4 t^2/U$ for a system with two side-coupled QDs. Other parameters are as in Fig. \ref{Aw-2QD-tx}. The lines represent the behavior of the Kondo temperature in the two correlated regimes. The dashed style line is the Kondo scale for a ferromagnetic model. The solid style line is a fit using the formula for the second stage Kondo temperature [see Eq. (\ref{eq:tksecond})] with fitting parameters $c_1=0.9$, $c_2=3.8$. }
\label{logTk-Aw-2QD-tx}
\end{figure}
The behavior of the low energy scale $k_BT_0$ as a function of $J=4t^2/U$ confirms this crossover from two-stage underscreened to ferromagnetic Kondo effects for $J\sim k_B T_K^0$. Fitting the numerical $A_0(\omega\to0)$ using Eq. (\ref{singAlow}) we obtained $T_0$ as a function of $J$ which we plot in Fig. \ref{logTk-Aw-2QD-tx}. The dependence of $T_0$ on $J$ for $J\ll k_BT_K^0$ is consistent with an underscreened Kondo regime $T_0\sim T_K^\star$ [see Eq. (\ref{eq:tksecond})]. For $J\gg k_BT_K^0$ the temperature scale $T_0$ is consistent with a ferromagnetic Kondo regime with a coupling $J_K$ given by Eq. (\ref{Jk_N}). 

\begin{figure}[tbp]
\includegraphics[width=\columnwidth]{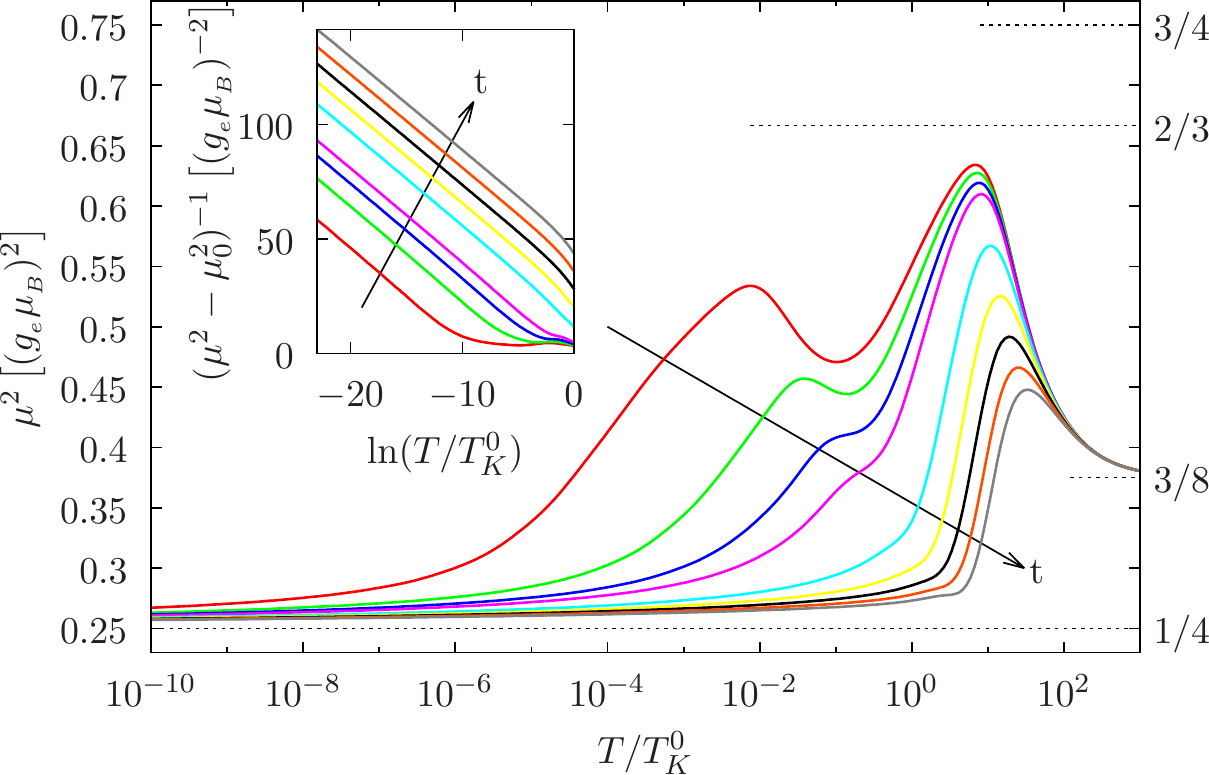}
\caption{(Color online) Magnetic moment squared of a triple QD device as a function of the temperature for different values of the interdot hopping $t$. All parameters are as in Fig. \ref{Aw-2QD-tx}, except for the $t=0$ curve which is not shown here. In the two stage Kondo regime ($4t^2/U\ll k_B T_K^0$) there is a two peak structure, while in the ferromagnetic Kondo regime ($4t^2/U\gg k_B T_K^0$) there is a single peak. The expected values of $\mu^2$ in four limiting cases (see text) are indicated on the right axis. In the whole range of values of $t$, the low temperature behavior is logarithmic as shown in the inset.}
\label{susN2}
\end{figure}

The underscreened and ferromagnetic Kondo regimes can also be identified through the behavior of the magnetic susceptibility contribution $\chi_{QD}$ of the QD array as a function of the temperature. 
The magnetic susceptibility $\chi_{QD}$ can be calculated assuming a coupling term:
\begin{equation}
    H_B= \boldsymbol{\mu} \cdot \mathbf{B}
\end{equation}
between the magnetic moment of the QD array $\boldsymbol{\mu}=g_e\mu_B \sum_\ell \mathbf{S_\ell}$ and an external magnetic field  $\mathbf{B}$.  For a free total spin $L$, the temperature dependence of the magnetic susceptibility follows a Curie law:
\begin{equation}
    \chi_{QD} = \frac{C}{T}, 
    \label{eq:Curie}
\end{equation}
where $C = (g_e \mu_B)^2 L(L+1)/3$ is the Curie constant. In the general case, the spin in the QD is not free and several spin multiplets may be relevant at a given temperature. It is nevertheless useful to define a temperature dependent magnetic moment squared $\mu^2(T)=\chi_{QD} T$, to analyze the screening processes.
The QD array contribution to the magnetic susceptibility is calculated, using the NRG, from the fluctuations of the magnetization at zero field subtracting the contribution from the electrodes \cite{wilson1975kondo}.
In Fig. \ref{susN2} we plot $\mu^2$ as a function of temperature.  In the high temperature regime ($k_B T  \gg U,\,t,\, \Delta$) all states of the QD array are equally probable which leads to $\mu^2\sim 3/8(g_e\mu_B)^2$, with each QD contributing $\sim 1/8(g_e\mu_B)^2$ to the magnetic moment. 
For $J\ll k_BT_K^0$, $\mu^2(T)$ has two peaks as a function of the temperature. The high temperature peak at $T\sim U/k_B$ is associated with the formation of a spin-$1/2$ magnetic moment on each of the QDs which for three isolated QDs
would lead to an increase in $\mu^2$ up to $3/4(g_e\mu_B)^2$. 
The Kondo screening of the magnetic moment of the central QD leads to a decrease in $\mu^2$ as the temperature is lowered which for $J\to0$ would reach a plateau at $1/2(g_e\mu_B)^2$ associated with the two remaining uncoupled magnetic moments on the side-coupled QDs. The low temperature peak in $\mu^2$ is due to the coupling of the magnetic moments of the side-coupled QD (through the central QD) to form a triplet state with spin $S=1$ [the magnetic moment squared associated with a spin 1 is $ 2/3(g_e\mu_B)^2$]. For lower temperatures the spin triplet is partially screened by the Kondo quasiparticles at the central QD leading to a residual magnetic moment $\mu_0^2=1/4(g_e\mu_B)^2$ as $T\to 0$, due to an asymptotically free spin-1/2. 

As $J$ increases ($t$ increases) the two peaks in $\mu^2$ decrease their amplitude and for $J\gtrsim k_B T_K^0$ the low temperature peak disappears. In the latter regime, the magnetic moments of the QDs couple to form a spin 1/2 at a temperature $T\sim J/k_B$ and the two screening processes of the $J\ll k_B T_K^0$ regime do not take place. The spin $1/2$, which is coupled ferromagnetically to the electrodes, decouples asymptotically as the temperature is lowered \cite{anderson1970poor} leading to a zero temperature magnetic moment $\mu_0^2=1/4(g_e\mu_B)^2$. 

In the whole range of values of $t\neq 0$ explored the low temperature magnetic moment presents a singular behavior of the form 
\begin{equation}
    \mu^2 \sim \mu_0^2\left(1 -\frac{1}{\ln(T/\widetilde{T}_0)}\right),
	\label{eq:sussing}
\end{equation}
as expected from the Bethe ansatz solutions of the underscreened and ferromagnetic Kondo models \cite{tsvelick1983exact,furuya1982bethe,andrei1983solution}.
This is illustrated in the inset to Fig. \ref{susN2} plotting $(\mu^2-\mu_0^2)^{-1}$ and a function of $\ln(T)$. Fitting the $\mu^2$ data with the expression of Eq. (\ref{eq:sussing}) we find $\mu_0^2\simeq 1/4(g_e\mu_B)^2$ and a temperature scale $\widetilde{T}_0\sim T_0$. 


\subsection{ $N$ dependence of the low-energy properties}
To reduce the computational cost of the numerical calculations we consider in what follows a weak interdot coupling regime where we can decouple the degrees of freedom associated with the empty and double occupied states in the side-coupled QDs and describe the coupling between QDs using Eq. (\ref{MultiHeisenb}). We also focus, for simplicity, our analysis on the electron-hole symmetric situation and $J_\ell=J$. We have checked numerically that our conclusions hold for a wide range of parameters where these conditions are not met, breaking the electron-hole symmetry and the symmetry of the exchange couplings.

The analysis of the previous section can be extended for any number of side-coupled quantum dots. As it was shown in Sec. \ref{sec:proof}, for a weak coupling of the central QD to the electrodes $k_B T_K^0\ll J$, the low energy Hamiltonian of the system is a ferromagnetic Kondo model for a spin $S=(N-1)/2$. As we show below, in the opposite situation where $J \ll k_B T_K^0$ this is not the case. As the temperature is lowered below $T_K^0$ the magnetic moment of the central QD is screened, and at lower temperatures ($T<J/k_B$) a magnetic moment $S=N/2$ forms on the side-coupled QDs which is partially screened below a characteristic temperature $\sim T_K^\star\ll T_K^0$.

\begin{figure}[tbp]
\includegraphics[width=\columnwidth]{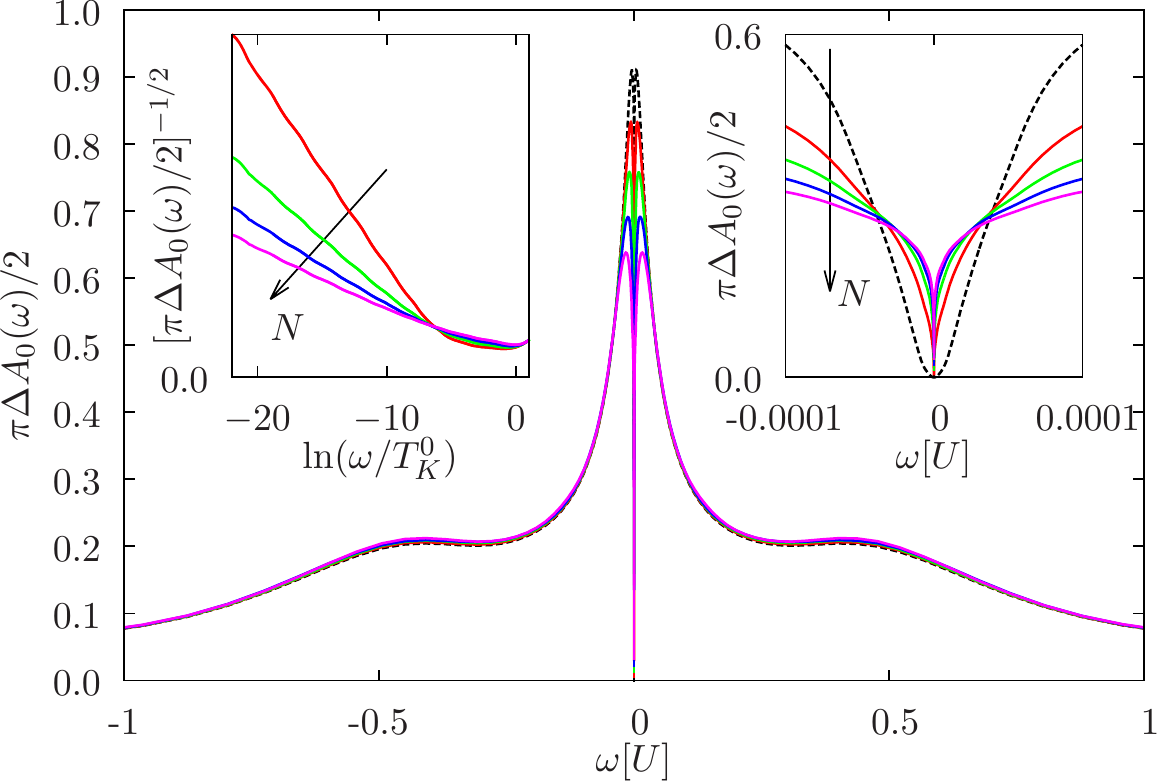}
\caption{(Color online) Spectral density of the central QD $A_0(\omega)$ for devices with $N=1,2,3,4$, and $5$ side-coupled QDs. The devices are in the two-stage Kondo regime $J=0.004$. The other parameters are $U_0=0.4$, $\varepsilon_0=-0.2$, and $\Delta=0.04\pi$. Left inset: plot of $A_0(\omega)^{-1/2}$ as a function of $\ln(\omega)$ to make clear the singular behavior of the spectral density at low energies for $N>1$. Right inset: low energy detail of the spectral density.}
\label{Aw-xQD-t002}
\end{figure}

Figure \ref{Aw-xQD-t002} presents the spectral density of the central QD for systems with up to $5$ side-coupled QDs. The systems are in the two-stage Kondo regime and $A_0(\omega)$ presents the same qualitative features for all values of $N$. Two charge transfer peaks associated with the spectrum of the central QD are located at $\omega=\epsilon_0, U+\epsilon_0$ and a central peak of width $T_K^0$  with a dip at the Fermi level of width $\sim k_BT_K^\star$. Due to the electron-hole symmetry considered, the spectral density vanishes exactly at the Fermi level which disagrees with the expectations from Fermi liquid theory for odd $N$. The ground state is, however, not a Fermi liquid for any $N>1$. 
The right inset in Fig. \ref{Aw-xQD-t002} shows that the quadratic behavior expected for a Fermi liquid is only obtained for $N=1$, for larger values of $N$ the spectral density vanishes logarithmically as described by Eq. (\ref{singAlow}) which is confirmed plotting $A_0(\omega)^{-1/2}$ as a function of $\ln(\omega)$ (see left inset in Fig. \ref{Aw-xQD-t002}).

\begin{figure}[tbp]
\includegraphics[width=\columnwidth]{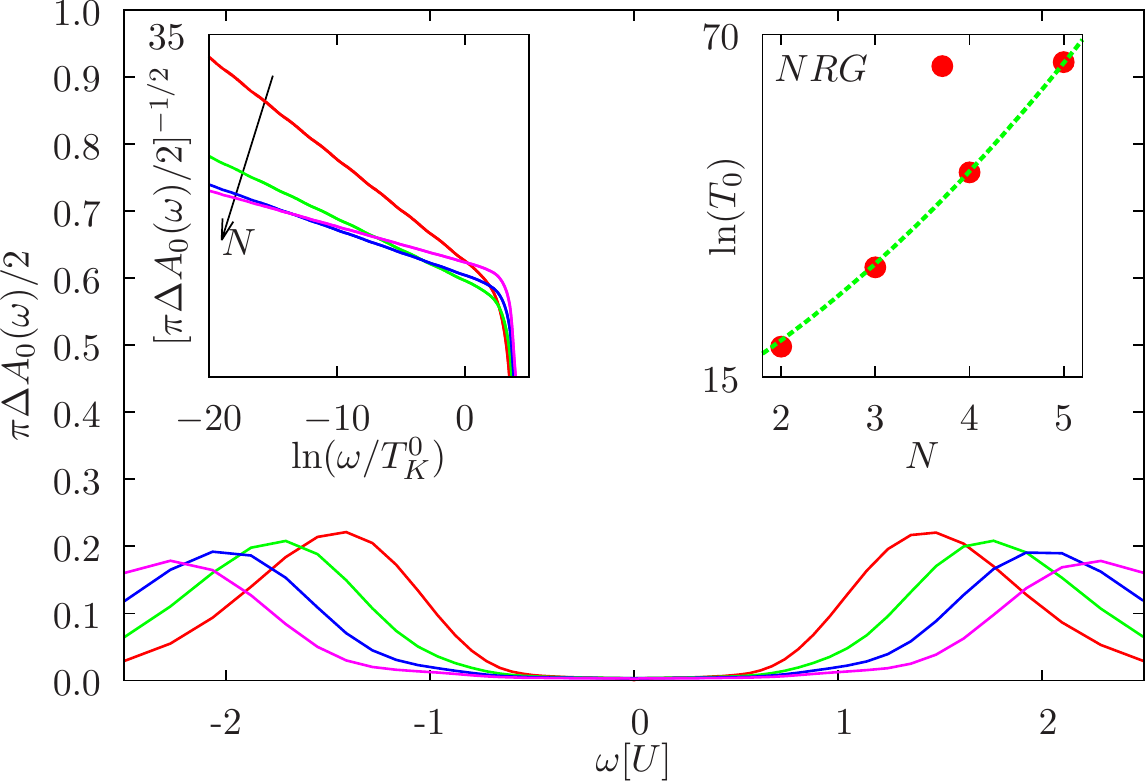}
\caption{(Color online) Same as in Fig. \ref{Aw-xQD-t002} for devices with $N=2,3,4$, and $5$ side-coupled QDs in the ferromagnetic Kondo regime, $J=0.4$. Left inset: plot of $A_0(\omega)^{-1/2}$ as a function of $\ln(\omega)$ to make clear the singular behavior of the spectral density at low energies for $N>1$. Right inset: $N$ dependence of the low energy scale $k_BT_0$. Numerical data (filled dots) and theory (dashed style line).}
\label{Aw-and-lnTk-xQD-t02}
\end{figure}

The spectral density in the ferromagnetic Kondo regime is presented in Fig. \ref{Aw-and-lnTk-xQD-t02} for systems with up to $5$ side coupled QDs. There is no Abrikosov-Suhl resonace in this case but a suppression of the density of states at low energies. For $\omega\to 0$ a logarithmic behavior is obtained (see left inset in Fig. \ref{Aw-and-lnTk-xQD-t02}) with a characteristic scale $T_0$ that is shown in the right inset of the figure. $T_0$ shows the expected behavior $T_0 \sim D/k_B \sqrt{|\rho_0 J_K|} \exp(-1/\rho_0 J_K)$, with $J_K$ given by Eq. (\ref{Jk_N}), as a function of $N$ which is presented as a dashed style line in the figure.

\begin{figure}[tbp]
\includegraphics[width=\columnwidth]{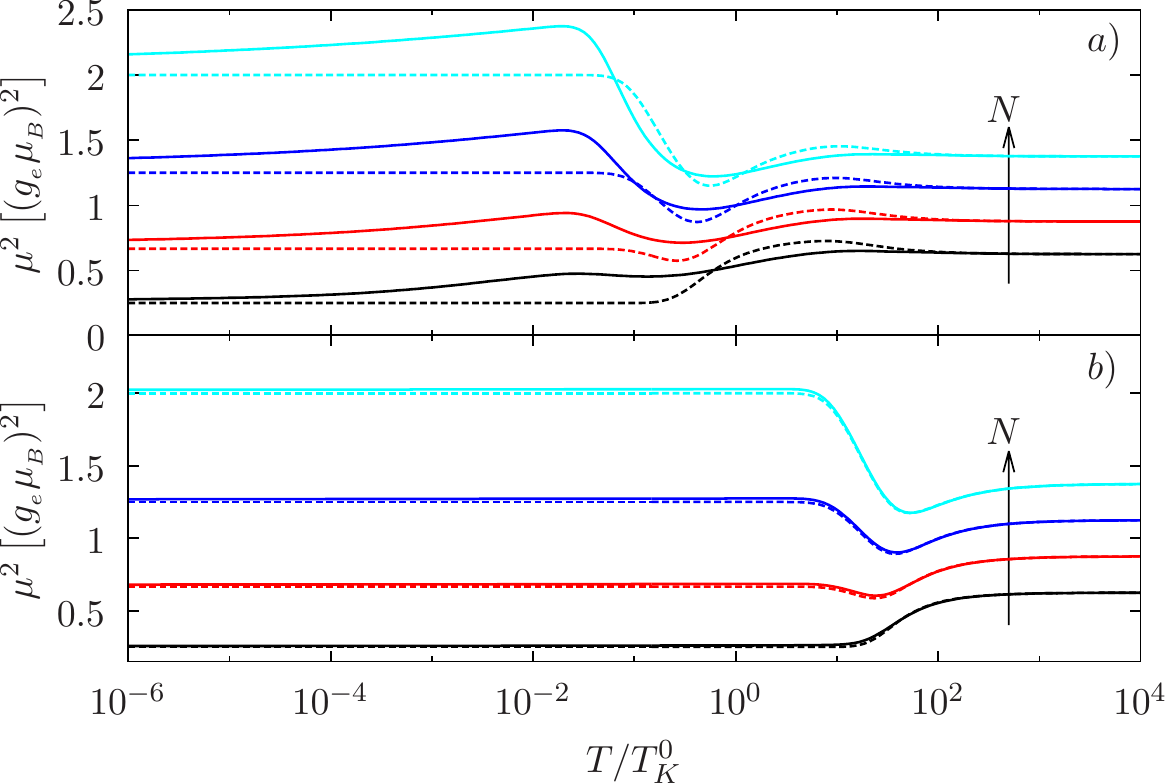}
\caption{(Color online)  Magnetic moment squared of the QD array for devices with $N=2,3,4$, and $5$ QDs side-coupled to the central QD. a) Two-stage Kondo regime ($J=0.004$). b) Ferromagnetic Kondo regime ($J=0.4$). The other parameters are as in Fig. \ref{Aw-xQD-t002}. 
The dashed lines correspond to the magnetic moment squared for the isolated QD array (V=0).}
\label{SusxQDsVp02}
\end{figure}

The low energy thermodynamic properties show a different behavior as a function of the temperature in the ferromagnetic and two-stage Kondo regimes. Figure \ref{SusxQDsVp02} presents the magnetic moment squared $\mu^2(T)$ of the QD array as a function of the temperature. We also present, as a reference, the values of $\mu^2(T)$ for the QDs decoupled from the electrodes ($V=0$). As expected in both the ferromagnetic Kondo and the underscreened Kondo regimes, the low and high temperature limits coincide with the isolated QD array results. In the $T\to 0$ limit both in the underscreened and in the ferromagnetic Kondo regimes the QD array has a magnetic moment $\mu_0^2=\mu^2(T=0)= S(S+1)/3=(N^2-1)/12$ as expected for a spin $S=(N-1)/2$ ferromagnetically coupled to the electrodes and to a spin $S=N/2$ underscreened by the conduction electrons. In the opposite limit $k_B T\gg V,J_\ell,U_0$ the QD array is effectively decoupled from the electrodes and from each other and there is a contribution to the magnetic moment of $N/4$ from the side-coupled QD and of $1/8$ from the central QD leading to $\mu^2(T\to\infty)= (2N+1)/8$. In both regimes, $\mu^2$ approaches the zero-temperature limit in a logarithmic way as described by Eq. (\ref{eq:sussing}) with $\mu_0^2\simeq (N^2-1)/12$ and $\widetilde{T}_0 \sim T_0$. 

In the ferromagnetic Kondo regime $\mu^2(T)$ deviates only slightly from the isolated QD array results. The coupling to the electrodes produces a small deviation which vanishes logarithmically as the temperature is lowered and the QD array effectively decouples from the electrodes. 
In the underscreened Kondo regime, however, the behavior of $\mu^2(T)$ is more complex and the two Kondo screening processes can be observed as the temperature is lowered. For $k_BT\lesssim U_0$ a magnetic moment forms in the central QD, leading to an increase in $\mu^2$, which at lower temperatures is screened by the conduction electrons. For thermal energies lower than $J$, the spins on the side coupled QD form a magnetic moment associated to a spin $S=N/2$ leading to a peak in $\mu^2$ that reaches $\sim N(N+1)/12$ for $k_BT\sim J$. The magnetic moment is partially screened leading to a logarithmic reduction of $\mu^2$ as $T\to0$.

\section{Summary and Conclusions} \label{sec:concl}
We have analyzed the properties of a $N+1$ QD dot device in a star configuration and showed that, as a function of the tunnel coupling $V$ between the central QD an the electrodes, there is a crossover between a ferromagnetic Kondo regime and a two-stage Kondo regime. In the weak coupling regime, the low energy properties are described by a ferromagnetic Kondo Hamiltonian for a spin $S=(N-1)/2$ impurity. In the strong coupling regime two Kondo effects take place successively as the temperature is lowered, a first stage were the spin-1/2 of the central QD is screened by a regular Kondo effect and a second stage were a spin $S=N/2$ in the side-coupled QDs is partially screened and reduced to  a spin $S=(N-1)/2$ in the ground state. In the full range of values of $V\neq 0$ the ground state of the system has spin $S=(N-1)/2$ and the low energy dynamic and thermodynamic properties have a singular behavior.

A numerical analysis of systems with up to 5 side-coupled QDs shows a singular behavior in the low energy spectral density of the central QD which determines the linear transport properties of the device and is accessible experimentally via spectroscopic measurements, in which one of the electrodes is weakly coupled to the central QD. We will present in a forthcoming publication the thermoelectric and magnetoelectric properties of the device that also show clear signatures of the singular low energy behavior of the device and allow to distinguish the different strongly correlated regimes. 

This QD devices would allow the observation of the so far elusive ferromagnetic Kondo effect for a spin $S\geq 1/2$ and the underscreened Kondo effect for a spin $S\geq 1$. Devices with more than five QDs may be difficult to construct therefore limiting the maximum value of the spin in the impurity models. The models described could be also applied to magnetic molecules coupled to electrodes or arrays of magnetic atoms on metallic surfaces \cite{ternes2015spin}.

\acknowledgments
This work was partially supported by CONICET PIP0832, SeCTyP-UNCuyo 06C347, and PICT 2012-1069.
\bibliographystyle{apsrev4-1}
\bibliography{references}

\end{document}